\documentclass[english]{article}

\usepackage{graphicx}
\usepackage{geometry}
\usepackage[T1]{fontenc}
\usepackage[latin9]{inputenc}
\usepackage{amsbsy}
\usepackage{caption}
\usepackage{subcaption}
\usepackage{setspace}
\usepackage{multirow}
\usepackage{hyperref}
\usepackage[authoryear]{natbib}
\usepackage{amsmath}
\usepackage{amssymb}
\usepackage{booktabs}

\newcommand{\RNum}[1]{\uppercase\expandafter{\romannumeral #1\relax}}
\newcommand{\squarebrackets}[1]{\left[#1 \right]}
\newcommand{\integral}[2]{\int_{#1}^{#2}} 
\newcommand{\parentheses}[1]{\left ( #1\right )}

\begin{document}
\title{Tuning diagonal scale matrices for HMC}

\author{Jimmy Huy Tran\footnote{Corresponding author, email: jimmy.tran@uis.no, Department of Mathematics and Physics, University of Stavanger, 4036 Stavanger, Norway} 
  \ and Tore Selland Kleppe\footnote{Department of Mathematics and Physics, University of Stavanger, 4036 Stavanger, Norway}}
\maketitle

\begin{abstract}
Three approaches for adaptively tuning diagonal scale matrices for HMC are discussed and compared. The common practice of scaling according to estimated marginal standard deviations is taken as a benchmark. Scaling according to the mean log-target gradient (ISG), and a scaling method targeting that the frequency of when the underlying Hamiltonian dynamics crosses the respective medians should be uniform across dimensions, are taken as alternatives. Numerical studies suggest that the ISG method leads in many cases to more efficient sampling than the benchmark, in particular in cases with strong correlations or non-linear dependencies. The ISG method is also easy to implement, computationally cheap and would be relatively simple to include in automatically tuned codes as an alternative to the benchmark practice. 

\end{abstract}



\doublespacing 
\section{Introduction} \label{intro}

For complex nonlinear Bayesian models, posterior analysis generally requires numerical methods, often in the form of Markov chain Monte Carlo (MCMC) methods \citep[see e.g.][ for an overview]{2208.00646}. Recently, Hamiltonian Monte Carlo (HMC) methods \citep[see e.g.][]{neal2011mcmc,bou-rabee_sanz-serna_2018} have proven particularly useful for such situations, in part as HMC methods scale very well when the dimension of the parameter space is large \citep[see e.g.][]{10.1214/21-AIHP1197}. 

There exist multiple variants of HMC-like algorithms, with the most prominent being the NUTS algorithm of \cite{hoffman2014no} which is implemented in the widely used Stan software \citep{JSSv076i01}. 
Common to these methods is that they involve a set of tuning parameters that may affect their performance strongly. In particular, the mass matrix in the kinetic energy term of the Hamiltonian (also referred to as the scale matrix for the remaining of the text after a certain transformation) plays a very important role in the performance of the sampling process \citep[see e.g.][]{neal2011mcmc}.

If the target distribution is approximately Gaussian, and somehow a good estimate of the covariance matrix $\boldsymbol{\Sigma}$ is available, \cite{neal2011mcmc} proposes to choose the mass matrix in the kinetic energy term to be $\boldsymbol{\Sigma}^{-1}$. 
The picture is less clear if the target density is highly non-Gaussian. This has led to the development of methods involving position-dependent scale matrices \citep{girolami_calderhead_11}. However, selecting such variable scale matrices for general models is highly non-trivial \citep{beatan_hessian,Kleppe2017,kleppe2023}.
Further, for the purposes of computational efficiency, one is in many cases restricted to work with diagonal scale matrices as the dimensionality of the problem at hand prevents full numerical linear algebra operations at each numerical integrator step. It is not obvious that simply taking a diagonal scale matrix to reflect the marginal variances of each sampled component would lead to efficient sampling in cases where non-trivial correlations are present.

Here, a set of different approaches for tuning (fixed) diagonal scale matrices are discussed and compared numerically. Particular emphasis is put on methods that could be included in general purpose HMC codes, with otherwise few assumptions with respect to the model at hand. The comparison is carried out within the recently proposed Generalized Randomized Hamiltonian Monte Carlo (GRHMC) framework \citep{bou2017randomized,kleppe2022connecting}. GRHMC processes are continuous time piecewise deterministic Markov process \citep{fearnhead2018piecewise} with the deterministic dynamics determined by the Hamilton's equations. The mass matrix tuning procedures presented here rely on numerical representations of continuous time dynamics, but similar developments could also be performed for more conventional (discrete time) HMC methods based on the Verlet leapfrog integrator. Therefore, the finds of this paper should also be relevant for such methods as the same underlying Hamiltonian dynamics underpin both methods.

Along with the more conventional scaling by marginal standard deviations (e.g. used as default by Stan \citep[][Section 15.1]{stan2018}), a variant of the integrated squared gradient (ISG) method proposed by \cite{kleppe2022connecting} is considered. The ISG offers two distinct interpretations, with the former being that for Gaussian targets, the ISG scales the target in accordance with the estimated target precision matrix diagonal elements. Secondly, the ISG may be interpreted as a technique for scaling the variables so that the underlying second order differential equation involves forces which are, on average, of a similar magnitude.

The derivation of the third technique considered does not rely on any Gaussian assumptions, but rather it exploits the Hamiltonian dynamics associated with the given target distribution at hand. The diagonal elements of the scale matrix can be thought of as parameters reflecting how fast the Hamiltonian dynamics explore the range of the corresponding position coordinates. 
This also implies that the \emph{median crossing times} of this coordinate, i.e. the time between two consecutive events where each coordinate crosses its corresponding median, will on average, be modulated by the scale matrix diagonal.
This motivates the idea of using median crossing times to tune the scale matrix. More specifically, the goal is to adaptively tune the scale matrix such that 
the mean median crossing times for all coordinates are similar to each other.

The reminder of the paper is laid out as follows. Section \ref{background} covers background on HMC methods and their scaling. Section \ref{approach tuning scale matrix} provides a detailed description of the different methods for tuning the scale matrix, Section \ref{simulation} provides a numerical comparison of their relative merit, and a further comparison based on Bayesian logistic regression applied to real data is carried out in Section \ref{bayesian logistic regression}. Finally, Section \ref{discussion} provides some discussion. Source code for the implementations and simulations is available at \url{https://github.com/jihut/tuning_scale_matrix_hmc}.

\section{Background and setup} \label{background}

\subsection{Standard Hamiltonian Monte Carlo} \label{standard hmc}
The aim is to sample from a continuous distribution with probability density function $\pi(\mathbf{q}) \propto c \Tilde{\pi}(\mathbf{q}),\;\mathbf{q} \in \mathbb{R}^d $, where $c$ is an unknown normalizing constant and $\Tilde{\pi}(\mathbf{q})$ is the density kernel of the distribution that may be evaluated. Common to Hamiltonian Monte Carlo methods (see e.g. \cite{neal2011mcmc}) is that the variable of interest $\mathbf{q}$ is taken to be the position vector in a Hamiltonian system, along with introducing an auxiliary momentum vector $\mathbf{p} \in \mathbb{R}^d$. The Hamiltonian of the system is given by: 
\begin{equation} \label{2.1}
    \mathcal{H}(\mathbf{z}) = \mathcal{H}(\mathbf{q}, \mathbf{p}) = - \log \Tilde{\pi}(\mathbf{q}) +  \frac{1}{2} \mathbf{p}^T \mathbf{M}^{-1} \mathbf{p}.
\end{equation}
Here, $\mathbf{z} = (\mathbf{q}, \mathbf{p}) \in \mathbb{R}^{2d}$ and $\mathbf{M} \in \mathbb{R}^{d \times d}$ is a symmetric and positive definite matrix known as the mass matrix. 
The Boltzmann-Gibbs distribution associated with (\ref{2.1}) is obtained as 
\begin{equation} \label{2.2}
    P(\mathbf{z}) \propto \exp(-\mathcal{H}(\mathbf{z})). 
\end{equation}
The (time-)evolution of the system is governed by the Hamilton's equations, given as 
\begin{align} 
    \Dot{\mathbf{q}}(t) &= \nabla_{\mathbf{p}} \mathcal{H}(\mathbf{z}(t)) = \mathbf{M}^{-1} \mathbf{p}, \label{2.3} \\
    \Dot{\mathbf{p}}(t) &= -\nabla_{\mathbf{q}} \mathcal{H}(\mathbf{z}(t)) = \nabla_{\mathbf{q}} \log \Tilde{\pi}(\mathbf{q}) \label{2.4},
\end{align}
where $\nabla_{\mathbf{p}} \mathcal{H}$ and $\nabla_{\mathbf{q}} \mathcal{H}$ denote the gradient of $\mathcal{H}$ with respect to $\mathbf{p}$ and $\mathbf{q}$, respectively, and dots indicate derivatives with respect to time. 

The solutions of (\ref{2.3}) and (\ref{2.4}) leave the Boltzmann-Gibbs distribution (\ref{2.2}) invariant \citep[see e.g.][]{neal2011mcmc}. I.e., suppose $\mathbf{z}(t) \sim P$, then for any time increment $\tau$, $\mathbf{z}(t + \tau) \sim P$ provided $\mathbf{z}(t + r),\;r\in(0,\tau)$ solves (\ref{2.3},\ref{2.4}). This property is crucial for applying Hamiltonian dynamics in a Monte Carlo simulation setting. Under (\ref{2.2}), $\mathbf{q}$ and $\mathbf{p}$ are independent with the target distribution corresponding to the marginal distribution of $\mathbf{q}$. Further, $\mathbf{p} \sim N(\mathbf{0_d}, \mathbf{M})$ with $\mathbf{0_d}$ denoting the $d$-dimensional zero vector.

An ideal HMC-like algorithm would involve alternating between propagating the Hamiltonian dynamics for given periods of time, and refreshing the momentum by simulating $ \mathbf p \sim N(\mathbf{0_d}, \mathbf{M})$ \citep[see e.g.][for a discussion of different momentum refresh strategies]{pmlr-v151-hoffman22a}. In the most basic HMC algorithm, the leapfrog method \citep[see e.g.][]{Leimkuhler:2004} is used in order to simulate the Hamiltonian dynamics, as it is not possible to solve the Hamilton's equations analytically in most cases. Since this integrator is time-reversible and volume preserving, it can be shown that the numerical error from the leapfrog method can be exactly corrected using a Metropolis correction where the acceptance probability only involves the Hamiltonian at the two endpoints of a simulated Hamiltonian trajectory (\cite{neal2011mcmc}). 

\subsection{Target standardization} \label{target standardization}
As mentioned in the introduction, the mass matrix $\mathbf M$ should reflect the scaling properties of the target distribution in order to obtain the best possible performance. Here, rather than to work with a non-identity mass matrix $\mathbf M$, an affine re-scaling of the target distribution is introduced. This re-scaling allows the mass matrix of the Hamiltonian system that is actually simulated to be the identity matrix. The re-scaling approach is convenient, in particular in light of the numerical solution of (\ref{2.3},\ref{2.4}), as both position- and momentum coordinates have roughly unit magnitudes when appropriate re-scalings are employed. 

More specifically, the canonical transformation \citep[see e.g.][]{goldstein2002classical} $\mathbf{q} = \mathbf{m} + \mathbf{S} \mathbf{\Bar{\mathbf{q}}}$, $\mathbf p=\mathbf S^{-T} \mathbf{\Bar{\mathbf{p}}}$ is introduced, where $\mathbf{m} = (m_1, m_2, \dots, m_d)^T \in \mathbb{R}^d$ is interpreted as the center vector and $\mathbf{S} \in \mathbb{R}^{d \times d}$ as a scale matrix so that $\mathbf S \mathbf S^T = \mathbf M^{-1}$. The transformation leads to the standardized target density 
\begin{equation} \label{2.5}
    \Bar{\pi}(\Bar{\mathbf{q}}) = \pi(\mathbf{m} + \mathbf{S}\mathbf{\Bar{\mathbf{q}}})  \lvert \mathbf{S} \rvert \propto \Tilde{\pi}(\mathbf{m} + \mathbf{S}\mathbf{\Bar{\mathbf{q}}}).
\end{equation}
Further, the standardized momentum variable  $\mathbf{\Bar{\mathbf{p}}}$ will have identity mass matrix, which leads to the Hamiltonian
\begin{equation} \label{2.6}
    \mathcal{H}(\mathbf{\Bar{z}}) = \mathcal{H}(\mathbf{\Bar{\mathbf{q}}}, \mathbf{\Bar{\mathbf{p}}}) = - \log{\Bar{\pi}}(\mathbf{\Bar{\mathbf{q}}}) +  \frac{1}{2} \mathbf{\Bar{\mathbf{p}}}^T \mathbf{\Bar{\mathbf{p}}}.
\end{equation}
Accordingly, the Hamilton's equations can be specified as 
\begin{align} 
    \mathbf{\Dot{\Bar{q}}}(t) &= \mathbf{\Bar{\mathbf{p}}}(t) \label{2.7}, \\
    \mathbf{\Dot{\Bar{p}}}(t) &= \nabla_{\mathbf{\Bar{\mathbf{q}}}} \log \Bar{\pi}(\Bar{\mathbf{q}}) =  \mathbf{S}\squarebrackets{\nabla_{\mathbf{q}} \log \Tilde{\pi} (\mathbf{q})} \label{2.8}, \quad \mathbf{q} = \mathbf{m} + \mathbf{S} \mathbf{\Bar{q}}.
\end{align}
For computational considerations, $\mathbf{S}$ is taken to be a diagonal matrix, and the methods to select such a diagonal matrix will be discussed shortly. Depending on the approach used in order to find the diagonal elements of $\mathbf{S}$, $\mathbf{m}$ is taken to be either the (estimated) median or mean of $\pi$. Further, note that the original parameterization $\mathbf q$ is computed in (\ref{2.8}), and therefore obtaining samples targeting $\pi(\mathbf q)$ generally comes without any additional computational cost.


\subsection{Generalized randomized Hamiltonian Monte Carlo process} \label{grhmc}
Rather than to work with a conventional (discrete time) HMC MCMC algorithm, sampling is performed here using numerical Generalized randomized Hamiltonian Monte Carlo (GRHMC) processes \citep{bou2017randomized,kleppe2022connecting,kleppe2023,2311.14492}. Numerical GRHMC processes are based on simulating the underlying Hamiltonian dynamics using high quality adaptive numerical integrators \citep[see e.g.][]{10.5555/153158} without Metropolis correction. This practice introduces arbitrarily small biases, which are controlled by adjusting the integrator error tolerances \citep{kleppe2022connecting} (at the cost of more computing). As the same (error tolerance) tuning parameters may be used throughout, the application of numerical GRHMC facilitates a clean comparison of the different methods for choosing the scale matrix $\mathbf S$. Further, the scale matrix tuning methods discussed allow for easy implementation in the continuous time/adaptive integration setting. Still, similar methods could certainly be developed also for discrete time HMC based on symplectic integrators (with suitable interpolation formulas) and the finds regarding $\mathbf{S}$-tuning methods also carry over to discrete time HMC methods.

GRHMC processes are continuous time piecewise deterministic Markov process (PDMP) \citep[see e.g.][]{davis1984piecewise,fearnhead2018piecewise} with Hamiltonian dynamics (\ref{2.7},\ref{2.8}) above as the deterministic dynamics between events. Events occur as a Poisson process with a specified nonnegative event rate $\lambda(\mathbf{\Bar{z}}(t))$ that may depend on the current state of the process. At an event time denoted as $t^*$, the process is updated according to a transition kernel $Q(\cdot  \mid \mathbf{\Bar{z}}(t^*-))$, where $\mathbf{\Bar{z}}(t^* - )$ is the state of the process immediately before the event time $t^*$.

\cite{kleppe2022connecting} provides a transition kernel $Q( \cdot \mid \mathbf{\Bar{z}}(t^*-))$ which ensures that $P(\mathbf{\Bar{z}}) \propto \exp(-\mathcal H (\mathbf{\Bar{z}}))$ is the stationary distribution of  $\mathbf{\Bar{z}}(t)$ for a general event rate function $\lambda(\mathbf{\Bar{z}}(t))$. This $Q$ refreshes the momentum variable $\mathbf{\Bar{p}}$ while keeping $\mathbf{\Bar{q}}$ fixed in a similar manner as momentum refreshes in conventional discrete time HMC. When the event rate is independent of $\mathbf{\Bar{p}}$, preservation of the marginal $N(\mathbf{0}, \mathbf{I}_d)$ is sufficient for the update of $\mathbf{\Bar{p}}$. 

Combining all these components with a numerical ODE solver, one can simulate a trajectory of the process $\mathbf{\Bar{z}}(t),\; t \in \squarebrackets{0, T}$, where $T$ is the (time-)length of the GRHMC trajectory. In this work, the LSODAR solver of ODEPACK \citep{hindmarsh1982odepack} available in R through the \texttt{deSolve}-package \citep{Soetaert2010-JSS-deSolve} is used throughout as this solver also has capabilities for handling the momentum refresh events. Discrete time samples targeting the invariant distribution $P(\mathbf{\Bar{z}})$ can be obtained by specifying a sample spacing $\Delta > 0$ and letting $\mathbf{\Bar{z}}_{(i)} = \mathbf{\Bar{z}}(t_\text{burn} + \Delta i),\;i=0,\dots,\;\Delta i<T - t_\text{burn}$. Here, $t_\text{burn} \in \squarebrackets{0, T}$ denotes the burn-in period. Samples targeting $\pi(\mathbf q)$ is acquired by $\mathbf{q}_{(i)} = \mathbf{m} + \mathbf{S} \mathbf{\Bar{q}}_{(i)}$.

It is common that expectations of functions $\mathcal{M}$ with respect to the target distribution $\pi$, i.e. $E_\pi(\mathcal{M}(\mathbf{q}))$, are required. Such expectations may naturally be estimated using $N$ discrete samples $\sum_i \mathcal{M}(\mathbf{q}_{(i)}) / N = (\sum_i \mathcal{M}(\mathbf{m} + \mathbf{S} \mathbf{\Bar{q}}_{(i)}))/N$. However, due to the continuous time nature of the GRHMC processes, one may exploit the whole (post burn-in) trajectory for such estimation by:
\begin{equation}
    \frac{1}{T - t_\text{burn}}\integral{t_\text{burn}}{T} \mathcal{M}(\mathbf{m} + \mathbf{S} \mathbf{\Bar{q}}(t)) dt.\label{eq:M_int}
\end{equation}
See \cite{kleppe2022connecting} for a discussion of the relative merits of estimation of $E_\pi(\mathcal{M}(\mathbf{q}))$ using either discrete time samples or continuous time integrals (\ref{eq:M_int}). Note that integrals such as (\ref{eq:M_int}) may be easily approximated within the ODE solver. To see this, define
\begin{equation} \label{2.9}
    \mathbf{r}(t) = \integral{0}{t} \mathcal{M}(\mathbf{q}(u)) \, du. 
\end{equation}
Combining Hamilton's equations (\ref{2.7},\ref{2.8}) and $\Dot{\mathbf{r}}(t) = \mathcal{M}(\mathbf{q}(t))$, the system of first order differential equations
\begin{equation} \label{2.10}
    \begin{bmatrix}
        \Dot{\mathbf{\Bar{\mathbf{q}}}}(t) \\
        \Dot{\mathbf{\Bar{\mathbf{p}}}}(t) \\
        \Dot{\mathbf{r}}(t)
    \end{bmatrix} = 
    \begin{bmatrix}
        \mathbf{\Bar{\mathbf{p}}}(t) \\
        \mathbf{S}\squarebrackets{\nabla_{\mathbf{q}} \log \Tilde{\pi} (\mathbf{q})} \\
        \mathcal{M}(\mathbf{q}(t))
    \end{bmatrix},
    \quad \text{ where }
    \mathbf{q}(t) = \mathbf{m} + \mathbf{S} \mathbf{\Bar{q}}(t),
\end{equation}
may be used both for simulating (the between events) GRHMC trajectory and estimating time-integrated averages (\ref{eq:M_int}).
In particular, time-integrals of $\lambda(\mathbf{\Bar{z}}(t))$ required for simulating the event times \citep[see e.g.][Section 2.1]{fearnhead2018piecewise} are obtained by including $\lambda$ in $\mathcal M$.  

\section{Approaches for tuning the scale matrix} \label{approach tuning scale matrix}

\begin{figure}
     \centering
     \begin{subfigure}[b]{0.49\textwidth}
         \centering
         \includegraphics[width=\textwidth]{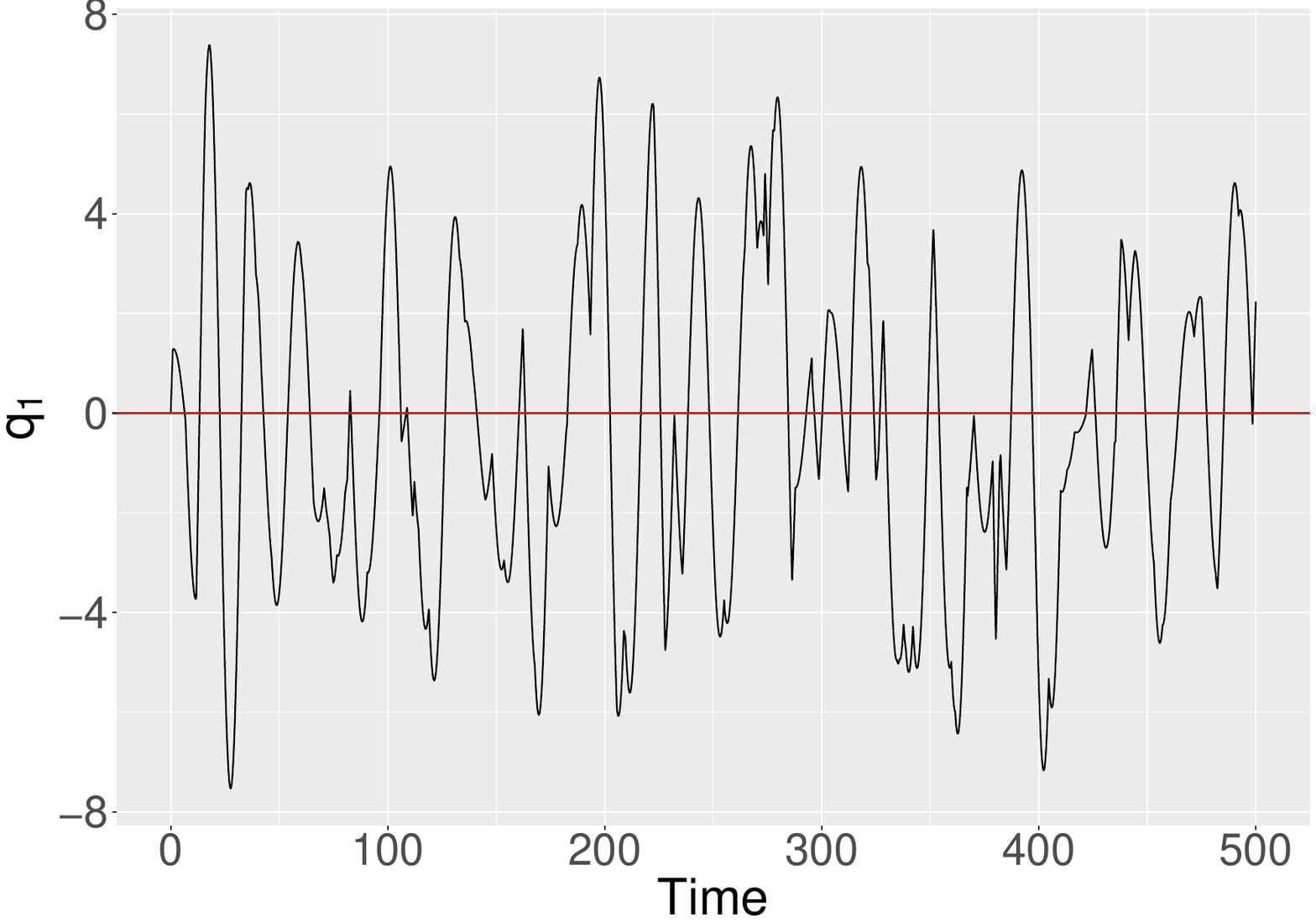}
         \caption{First coordinate with $\mu_1 = 0$ and $\sigma_1^2 = 10$.}
         \label{fig 1a}
     \end{subfigure}
     \hfill
     \begin{subfigure}[b]{0.49\textwidth}
         \centering
         \includegraphics[width=\textwidth]{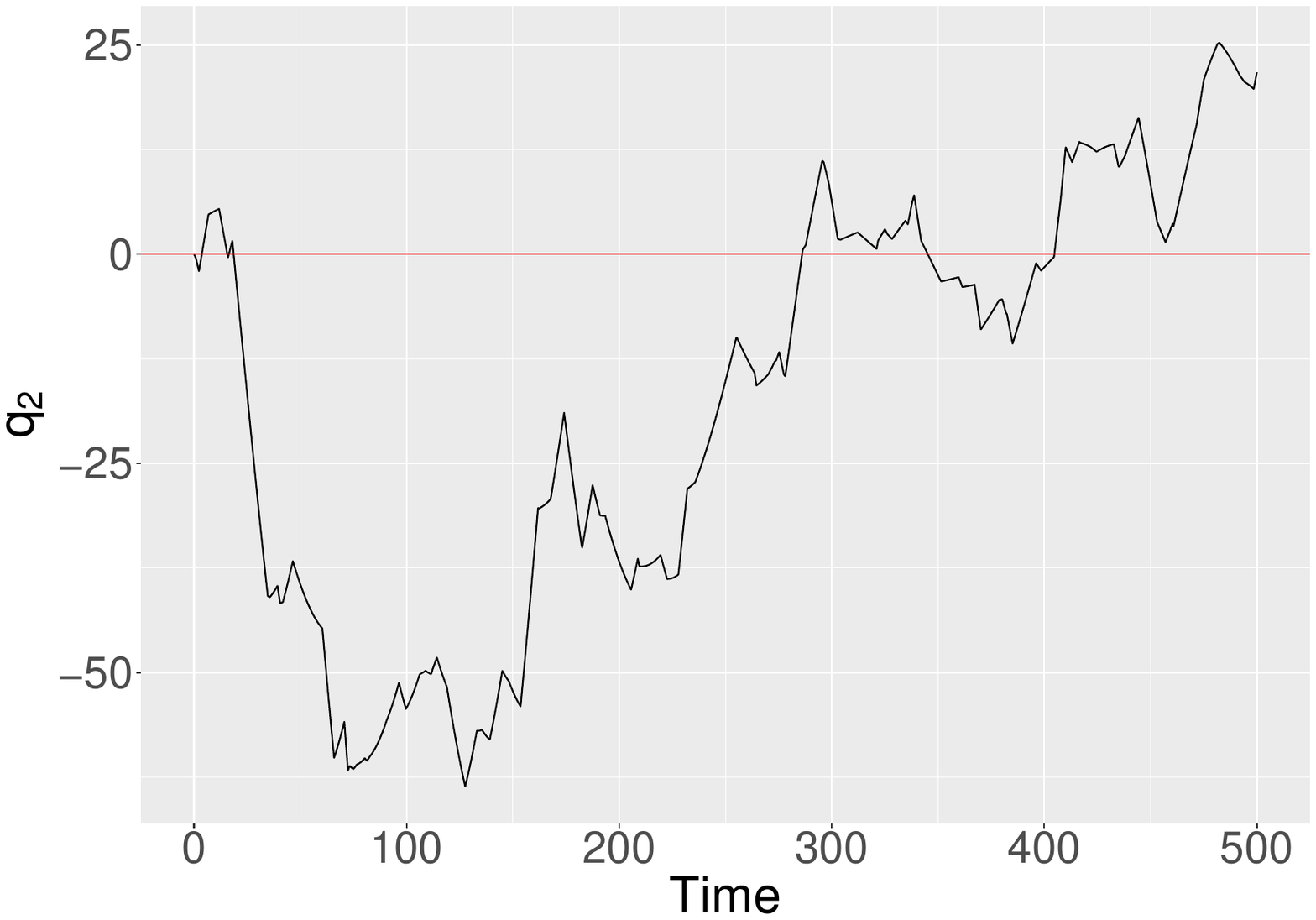}
         \caption{Second coordinate with $\mu_2 = 0$ and $\sigma_2^2 = 1000$.}
         \label{fig 1b}
     \end{subfigure}
     \caption{A numerical GRHMC trajectory targeting (\ref{eq:toy_example}) with $\mathbf S=\mathbf I_2$, $\lambda = 0.3$ and $T = 500$. Here, the red lines indicate the marginal mean/median of each coordinate.}
     \label{fig 1}
\end{figure}

So far, the diagonal elements of $\mathbf{S}$ have been left unspecified. The choice of scale matrix $\mathbf{S}$ can have a substantial impact on simulation performance, and systematic way to select appropriate values is further pursued here. 

Consider first a motivating example involving a bivariate normal distribution 
\begin{equation}
    \mathbf{q} \sim N(\mathbf{0}_2, \mathbf{V}), \text{ where } \mathbf{V} = 
\begin{bmatrix}
    10 & 5 \\
    5 & 1000
\end{bmatrix}.\label{eq:toy_example}
\end{equation}
Suppose now that the scale matrix is set to identity. Under the Boltzmann-Gibbs distribution associated with (\ref{2.6}), the velocity (which is equal to the momentum due to the identity mass in (\ref{2.6})) in each dimension has a standard Gaussian distribution. Sloppily speaking, it follows that the ground covered per unit of time is roughly the same in each dimension/direction.
One would therefore naturally expect that the relevant regions of the support of the coordinate with the smaller variance will be traversed many times during the time period it takes to traverse a single standard deviation of the second coordinate.
Figure \ref{fig 1} shows the two position coordinates of a GRHMC simulation targeting (\ref{eq:toy_example}) over 500 time units. The behaviors of the two marginal trajectories are very much in accordance with the discussion above.

\begin{figure}
     \centering
     \begin{subfigure}[b]{0.49\textwidth}
         \centering
         \includegraphics[width=\textwidth]{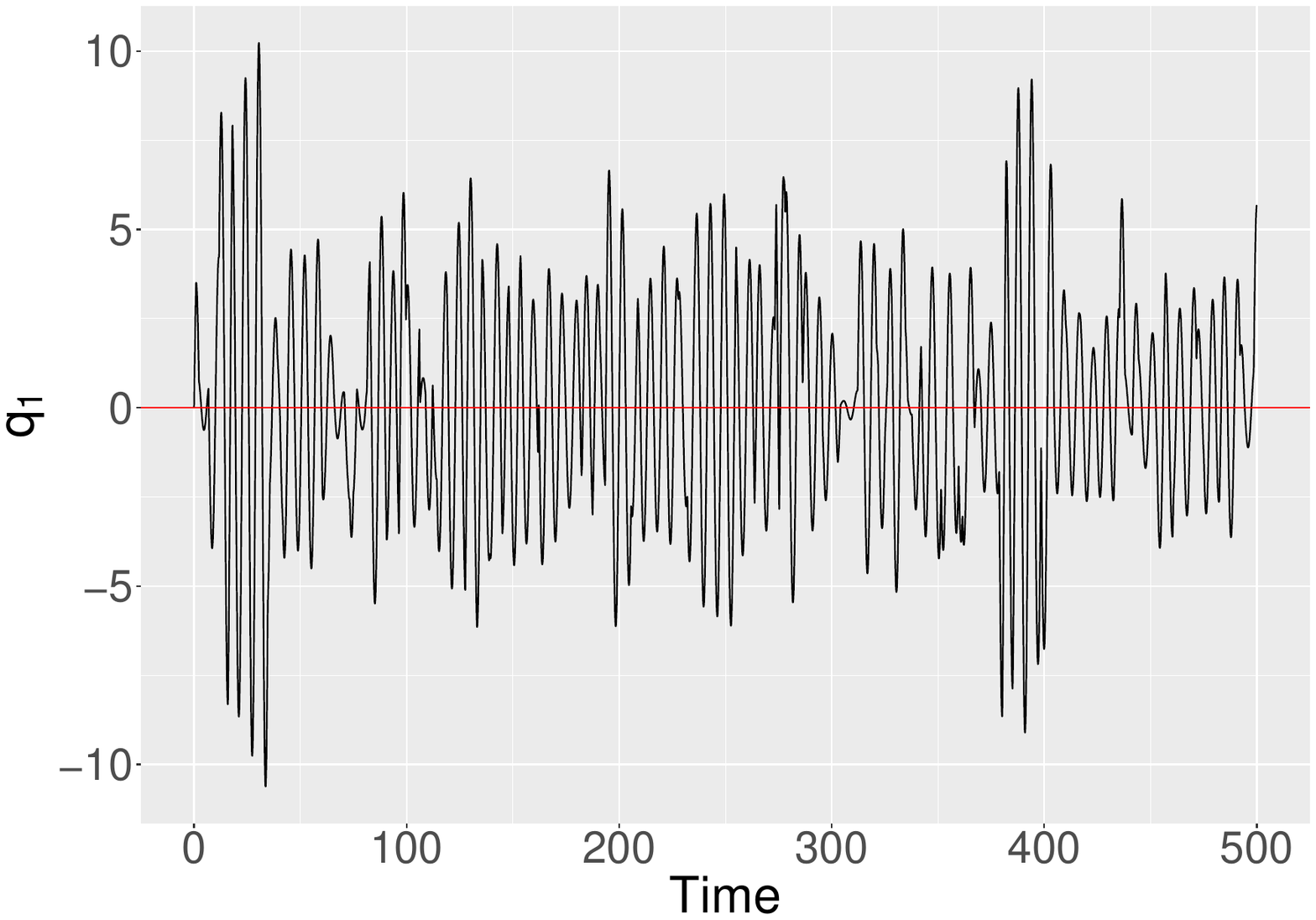}
         \caption{First coordinate with $\mu_1 = 0$ and $\sigma_1^2 = 10$.}
         \label{fig 2a}
     \end{subfigure}
     \hfill
     \begin{subfigure}[b]{0.49\textwidth}
         \centering
         \includegraphics[width=\textwidth]{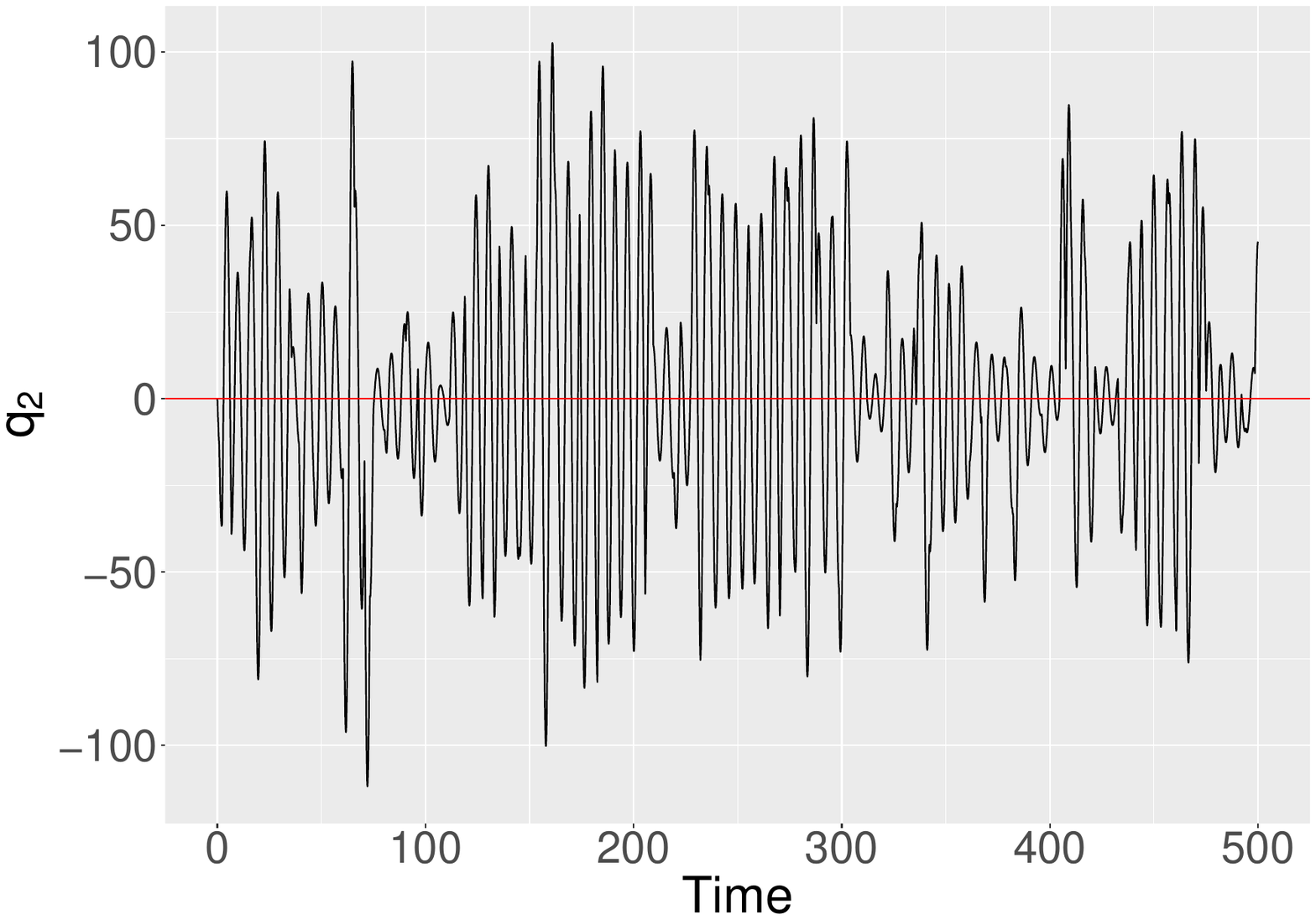}
         \caption{Second coordinate with $\mu_2 = 0$ and $\sigma_2^2 = 1000$.}
         \label{fig 2b}
     \end{subfigure}
     \caption{A numerical GRHMC trajectory targeting (\ref{eq:toy_example}) with $\mathbf S=\text{diag}(\sigma_1, \sigma_2)$, $\lambda = 0.3$ and $T = 500$. Here, the red lines indicate the marginal mean/median of each coordinate.}
     \label{fig 2}
\end{figure}

Figure \ref{fig 2} presents results obtained when changing the scaling matrix $\mathbf S$ in line with the marginal standard deviations of (\ref{eq:toy_example}), i.e. $\mathbf S=\text{diag}(\sqrt{10},\sqrt{1000})$. This change leads to simulations where the GRHMC trajectory covers the relevant parts of the support at approximately the same frequency. However, for non-Gaussian targets, or even Gaussian targets with non-trivial correlations, it is not clear that diagonal re-scaling using standard deviations is the optimal approach.

The reminder of this section presents three approaches that can be used to obtain values different from unity for the diagonal elements of $\mathbf{S}$, where each method optimizes $\mathbf{S}$ according to different objectives. The first two are based on results for a Gaussian distribution while the last one is more general and is inspired by the behaviors in Figure \ref{fig 1}. 

\subsection{Variance integrated (VARI)} \label{VARI}
As mentioned before in the introduction, the results from \cite{neal2011mcmc} indicate that for an approximate Gaussian setting, a good choice of $\mathbf{M}^{-1}$ is the estimated covariance matrix $\boldsymbol{\Sigma}$ if this quantity is obtainable. In the setup of Section \ref{target standardization}, this would amount to setting $\mathbf S$ to be the lower triangular Cholesky factor of $\boldsymbol{\Sigma}$. However, when restricted to a diagonal $\mathbf S$ or $\mathbf M^{-1}$, common practice \citep[see e.g.][ and Figure 2 here]{stan2018} is to choose $\mathbf M^{-1}$ such that it contains the estimated marginal variances along the diagonal.  

In line with this conventional routine, we follow \cite{kleppe2022connecting} in estimating marginal variances using temporal averages (\ref{eq:M_int}). More specifically, suppose $t^* \in [0, t_\text{burn})$ is an event time in the burn-in period. Further, let $\mathbf{S} = \text{diag}(S_1, \dots, S_d)$. Then, 
\begin{equation} \label{2.11}
    S_j^2 = \frac{1}{t^*}\integral{0}{t^*} q^2_j(u)\, du - \squarebrackets{\frac{1}{t^*} \integral{0}{t^*} q_j(u)\, du }^2, \, j = 1, \dots, d,
\end{equation}
where the diagonal elements are updated at each event time $t^*$. From now on, this method will be referred to as VARI (variance integrated) similar to \cite{kleppe2022connecting}, and constitutes the current practice/benchmark method. 

The center vector $\mathbf{m} = (m_1, \dots, m_d)$ is updated at each event time $t^*$ by the temporal averages targeting the mean of the respective coordinates for VARI, i.e.
\begin{equation}
    m_j = \frac{1}{t^*} \integral{0}{t^*} q_j(u)\, du, \; j=1,\dots,d,
    \label{eq:mean_integral}
\end{equation}
which again is also updated at each event time $t^*$. 

Notice that when updates of either $\mathbf m$ or $\mathbf S$ occur, the modifications are done in such a manner that $\mathbf q$ is left unchanged. This practice is kept consistent throughout all of the methods considered in this paper and is accomplished by moving $\mathbf{\Bar{q}}$ to exactly account for the new values of $\mathbf m$ and $\mathbf S$ when updates occur. In other words, momentarily denote $\mathbf{m}^*$, $\mathbf{S}^*$ and $\mathbf{\Bar{q}}^*$ as the new center vector, scale matrix and transformed position variable respectively, whereas the corresponding quantities without *s are the old values. Then, $\mathbf{\Bar{q}}^*$ is updated concurrently with $\mathbf m$ and $\mathbf S$ as 
\begin{equation*}
    \mathbf{\Bar{q}}^* = (\mathbf{S}^*)^{-1} \mathbf{S} \mathbf{\Bar{q}} + (\mathbf{S}^*)^{-1}\parentheses{\mathbf{m} - \mathbf{m}^*}.
\end{equation*}

\subsection{Integrated squared gradients (ISG)} \label{ISG}
In the application of numerical methods for differential equations, it is common practice \citep[see e.g.][Chapter 17]{numrecipes2007} to introduce different re-scalings in order to make the set of ODEs under consideration better suited for numerical integration. Here, the second method for choosing $\mathbf S$ seeks to standardize the force term of the second set of ODEs in (\ref{2.10}) to have unit marginal second order moments \citep[see][for a similar development]{kleppe2022connecting}. To accomplish this in theory, the $j$-th squared diagonal element of $\mathbf S$ should be set equal to $E_\pi[(\partial/\partial q_j \;\log \pi(\mathbf q))^2]$. Note that the force term of the former set of ODEs in (\ref{2.10}) will already have unit covariance due to the identity mass in (\ref{2.6}). Further, note also that under suitable regularity conditions on $\pi$, $E_\pi[\partial/\partial q_j \;\log \pi(\mathbf q)]=0$ \citep[see e.g.][Proposition 1]{kleppe2023} so that the mentioned marginal second order moments typically coincide with corresponding marginal variances.

Interestingly, this choice of diagonal scaling also has a different interpretation for Gaussian target distributions. Specifically, if $\mathbf{q} \sim N(\boldsymbol{\mu}, \mathbf{V})$, then
\begin{equation*}
    E_\pi \parentheses{\squarebrackets{\nabla_{\mathbf{q}} \log \Tilde{\pi}(\mathbf{q})} \squarebrackets{\nabla_{\mathbf{q}} \log \Tilde{\pi}(\mathbf{q})}^T} = \mathbf{V}^{-1}.
\end{equation*}
Hence, the above approach would amount to setting the mass matrix diagonal equal to the diagonal of the precision matrix. Therefore, for a diagonal $\mathbf V$, the present approach and variance-based approach underpinning VARI would coincide. However, in presence of dependencies within $\mathbf q$, or more generally for non-Gaussian targets, the two methods will be distinct. 

To obtain an operational method, $\mathbf S$ is updated at each event time $t^*$ according to
\begin{equation} \label{2.12}
    1 / S_j^2 = \frac{1}{t^*} \integral{0}{t^*} \squarebrackets{\frac{\partial}{\partial q_j} \log \Tilde{\pi}(\mathbf{q}(u))}^2 \, du, \;j=1,\dots,d.
\end{equation}
Due to the appearance of (\ref{2.12}), this approach is from now on referred to as the integrated squared gradients method (ISG). For ISG, $\mathbf m$ is also taken to be an estimate of $E_\pi[\mathbf q]$ by using (\ref{eq:mean_integral}). Notice that the required gradient elements in (\ref{2.12}) are already computed in the numerical ODE implementation, and including (\ref{2.12}) in $\mathcal M$ therefore only constitute trivial additional computational cost.


\subsection{Median crossing times (MCT)} \label{median crossing}
As should be clear, the former two approaches have, at least to some degree, Gaussian assumptions underpinning their development. On the other hand, the last method considered takes as vantage point to synchronize the frequency of median crossing events across the different dimensions of $\mathbf q$. 

In Figure \ref{fig 1}, the occurrence of such events will arise far more frequently for the first coordinate (left panel) relative to the second (right panel). This motivates the idea of choosing the diagonal elements of $\mathbf{S}$ so that the two coordinates have similar frequency, as is observed in Figure \ref{fig 2}.
Equivalently, this is the same as choosing $\mathbf{S}$ in such a way that the median crossing time, i.e. the time between two consecutive median crossing events, is similar across the coordinates. Tuning the diagonal elements of $\mathbf{S}$ in this scenario is therefore comparable to the analogy of adjusting the velocity of each coordinate so that the time duration required for each coordinate to finish one cycle of exploring its range is similar for all coordinates. 

The approach is complicated by the fact that the mean MCT does not have a simple analytic expression in terms of $\mathbf S$. It is generally the case that the mean MCT of component $j$ will be decreasing as a function of $S_j$ since an increase in $S_j$ implies a reduction in the mass, which in turn leads to larger velocity if the momentum is fixed. In order to solve for $\mathbf S$ so that the mean MCT for each coordinate is close to some target time $\tilde T$, we resort to the
dual averaging technique of \cite{nesterov2009primal}.
Specifically, we use a modified version of the dual averaging of \cite{hoffman2014no} (who used the technique for tuning the time step size $\epsilon$ in the NUTS HMC procedure). The presented method will also be referred to as MCT from now on. 

Consider now a specific coordinate $j, \, j = 1, \dots, \, d$. Let the center vector $\mathbf{m}$ be the median vector (assumed momentarily, for simplicity, to be known). Let $\Tilde{t}_{k_j} \in (0, T], \, k_j = 1, 2, \dots$ denote the times $q_j(t)$ crosses its median $m_j$ for the $k_j$-th time. Note that this is equivalent to $\Bar{q}_j$ crossing zero, which will be the criterion used as the differential equations (\ref{2.7})-(\ref{2.8}) are in terms of the transformed variables. 

Further, let $\Tilde{T}_{k_j - 1} = \Tilde{t}_{k_j} - \Tilde{t}_{k_j - 1}, \, k_j = 2, 3, \dots$ be the recorded time between two consecutive median crossing events number $k_j - 1$ and recall that $\Tilde{T}$ is the desired average time between two such events. The main objective is to find $S_j$ such that when $S_j$ is fixed, the mean of all $\Tilde{T}_{k_j - 1}$ during the duration $T$ of the simulation is close to $\Tilde{T}$. To ensure that $S_j > 0$, the optimization is performed in $\log(S_j)$ instead of $S_j$. Following a similar strategy as the dual averaging scheme presented by \cite{hoffman2014no},  the sequence of random variables whose mean is to be brought to zero is
\begin{equation} \label{3.4}
    H_j^{(k_j)} = \Tilde{T} - \Tilde{T}_{k_j}. 
\end{equation}
Thus, 
\begin{equation} \label{3.5}
\begin{split}
    h_j(S_j) = \lim_{K \rightarrow \infty} \frac{1}{K} \sum_{k_{j}=1}^K E (H_j^{(k_j)} \mid \log(S_j)) = \Tilde{T} - \lim_{K \rightarrow \infty}\frac{1}{K} \sum_{k_{j}=1}^K E(\Tilde{T}_{k_j} \mid \log(S_j)) = 0
\end{split}
\end{equation}
defines the desired optimization problem. The update rule to find the parameter value for the next iteration after $k_j$ iterations is then given as follows: 
\begin{align} 
    \log(S_j^{(k_j+1)}) &= \mu_j - \frac{\sqrt{k_j}}{\gamma_j} \frac{1}{k_j + k_j^{(0)}} \sum_{i=1}^{k_j} H_j^{(i)}, \label{3.2} \\ 
    \log(\Bar{S}_j^{(k_j+1)}) &= \eta_{k_j} \log(S_j^{(k_j+1)}) + (1 - \eta_{k_j}) \log(\Bar{S}_j^{(k_j)}). \label{3.3}
\end{align} 
Here, $\mu_j$ is a prespecified value in which the estimates $S_j^{(k_j)}$ will be pushed towards, $\gamma_j > 0$ is a shrinkage parameter that determines the effect of this behavior, $k_j^{(0)} \geq 0$ is a chosen value for stabilization in the early stage of the optimization and $\eta_{k_j} = k_j^{-\kappa_j}$ is the learning rate at iteration $k_j$ with $\kappa_j \in (0.5, 1]$ (see e.g. \cite{robbins1951stochastic} and \cite{hoffman2014no} for further details). The subscript is to indicate that each coordinate can have its own set of parameters for the dual averaging approach. Since $h_j$ is a non-decreasing function, the average iterates (\ref{3.3}) of the parameter $\log(\Bar{S}_j^{(k_j)})$ from the dual averaging procedure will converge to a value that will also lead to $h_j(\log(\Bar{S}_j^{(k_j)}))$ converging to 0 (\cite{nesterov2009primal}).

In the procedure above, it is assumed that the median vector $\mathbf{m}$ is known. However, this is not true in practical applications and $\mathbf{m}$ must be estimated. In contrast to the mean used in VARI and ISG method, it is not straightforward to specify a form of (\ref{2.9}) to estimate the median continuously. Simply exchanging the median with mean is also not an option, as one may easily think of pathological examples where the Hamiltonian dynamics components only very infrequently pass their respective means. Rather, here, the median is estimated based on discrete samples. To alleviate the space complexity issues when storing all historical samples for all coordinates and estimating the median values by conventional methods, the procedure based on the $P^2$-algorithm by \cite{jain1985p2} is applied. Further details on how this algorithm is used for estimating $\mathbf{m}$ can be found in Appendix \ref{implementation median crossing}. Also, instead of updating $\mathbf{m}$ and $\mathbf{S}$ in (\ref{2.7}) and (\ref{2.8}) at momentum refresh events, this is rather done every single time a coordinate crosses its median as this is required in order for the dual averaging method to work in this setting. 

\section{Simulation experiments} \label{simulation}
In this section, simulation experiments and comparison of the three scale matrix tuning methods are presented. In order to cover a wide spectrum of situations, three categories of distributions are considered. The first part examines the performance of the methods for a set of Gaussian distributions. Application to non-linear target distributions, where it is expected that the results will differ to a larger degree across the methods, is then investigated. At last, the techniques are benchmarked using different types of the so-called funnel distribution (\cite{neal2003slice}). 

The absolute and relative tolerance of the LSODAR integrator \citep{hindmarsh1982odepack} used to solve the system of differential equations are set to $tol_r = tol_a = 10^{-6}$. This choice is overly strict \citep{kleppe2022connecting}, but is committed to ensure that biases introduced by the non-symplectic numerical methods will not affect the results. In practical applications, larger values of tolerances (leading to much higher performance) can safely be chosen. 

For each combination of method and distribution in the simulations below, 10 independent trajectories with time duration $t_\text{sample} = 100000$ and $N=50000$ are simulated ($t_\text{sample}$ corresponds to the duration of simulation after the burn-in period, see Appendix \ref{final setup} for more details). Each of the 10 replica has their own adaptation process. Finally, the end results are obtained by combining the samples from all 10 trajectories such that the final amount of discrete samples over all trajectories will be 500000. These will be the number of samples used in the calculation of effective sample sizes (ESS) \citep{geyer1992} based on the \texttt{rstan::monitor()} function of \texttt{rstan} \citep{stan2018,Vethari_ESS} for each given given combination of method and distribution. 

The burn-in period where the adaptation of $\mathbf{m}$, $\mathbf{S}$ and (constant) $\lambda$ occurs will follow the specified values in Appendix \ref{implementation median crossing} and \ref{final setup}, where both the implementation of MCT and the final setup used to produce the results in the remaining parts of the text are also explained in slightly more details. The implementation can be found at \url{https://github.com/jihut/tuning_scale_matrix_hmc}. 

\subsection{Normal distribution} \label{normal}
Many theoretical and analytical results are available for a multivariate normal distribution. This motivates choosing the normal distributions as a first example. For instance, consider a setting of (\ref{2.1}) and $\Tilde{\pi}$ corresponds to a standard multivariate normal distribution with mean vector (and therefore median vector) zero and identity matrix as the covariance matrix (e.g. $\mathbf m=\mathbf 0$, $\mathbf S=\mathbf I$). If $q_j(0) = 0$ such that the starting point is at the median of the distribution, it is known that an analytical solution of the position coordinate is $q_j(t) = r \cos(t), \, r > 0$. I.e. the solution will cross the median every $\pi$ units of time. This is the reason for the choice of $\pi$ as the desired median crossing times $\Tilde{T}$ (see also Appendix \ref{implementation median crossing}). 

\begin{table} 
\centering
\resizebox{\columnwidth}{!}{%
\begin{tabular}{ccccccccccc}
\toprule
 &  & \multicolumn{3}{c}{ESS} & \multicolumn{3}{c}{mean($\mathbf{S}$) (SD($\mathbf{S}$))} & \multicolumn{3}{c}{$\text{ESS} * 100000 $/ $N_{\text{ode}}$} \\
 \cmidrule(lr){3-5} \cmidrule(lr){6-8} \cmidrule(lr){9-11}
 & Method & VARI & ISG & MCT & VARI & ISG & MCT & VARI & ISG & MCT \\
 \cmidrule(lr){2-2} \cmidrule(lr){3-5} \cmidrule(lr){6-8} \cmidrule(lr){9-11}
Parameters & Position Coordinate &  &  &  &  &  &  \\
\midrule
\multirow{2}{*}{{G1: $\boldsymbol{\mu} = (1,2)^T$ \& $\boldsymbol{\Sigma} = 
\begin{bmatrix}
    4 & 0.5 \\
    0.5 & 9
\end{bmatrix}$}} & 1 & 736190 & 759191 & 739226 & 2.01 (0.067) & 2.02 (0.057) & 2.01 (0.030) & 3593 & 3759 & 3639 \\
 & 2 & 705132 & 706912 & 730685 & 2.99 (0.122) & 2.98 (0.099) & 3.01 (0.041) & 3442 & 3500 & 3597 \\
\cline{1-11}
\multirow{2}{*}{{G2: $\boldsymbol{\mu} = (0, 0)^T$ \& $\boldsymbol{\Sigma} = 
\begin{bmatrix}
    10 & 5 \\
    5 & 10^3
\end{bmatrix}$}} & 1 & 767228 & 738028 & 729982 &  3.19 (0.071) &  3.18 (0.107) &  3.16 (0.042) & 3785 & 3604 & 3586 \\
 & 2 & 711551 & 735771 & 705868 & 31.45 (0.997) & 31.77 (1.144) & 31.43 (0.535) & 3511 & 3593 & 3467 \\
\cline{1-11}
\multirow{2}{*}{{G3: $\boldsymbol{\mu} = (0, 0)^T$ \& $\boldsymbol{\Sigma} = 
\begin{bmatrix}
    1 & 0.95 \\
    0.95 & 1
\end{bmatrix}$}} & 1 & 293340 & 94345 & 280127 & 1.01 (0.033) & 0.31 (0.008) & 0.96 (0.069) & 379 & 449 & 384 \\
 & 2 & 293428 & 94380 & 279951 & 1.01 (0.034) & 0.31 (0.008) & 0.96 (0.065) & 379 & 449 & 384 \\
\cline{1-11}
\multirow{2}{*}{{G4: 10-dimensional standard normal distribution}} & max ESS & 932542 & 958789 & 916589 & 1.01 (0.017) & 1.01 (0.038) & 1.01 (0.014) & 4626 & 4788 & 4491 \\
 & min ESS & 845400 & 855352 & 873673 & 0.99 (0.023) & 1.00 (0.026) & 1.00 (0.008) & 4193 & 4271 & 4281 \\
\cline{1-11}
\bottomrule
\end{tabular}%
}
\caption{Results for a selection of normal distributions. $N_\text{ode}$ is the number of evaluations of the ODE system done by the numerical integrator in total over 10 independent trajectories. mean($\mathbf{S}$) and SD($\mathbf{S}$) correspond to the mean, and the standard deviation respectively, of the 10 values of the diagonal elements of $\mathbf{S}$ obtained from the 10 independent trajectories.}
\label{table 1}
\end{table}

Consider first the bivariate normal setting defined as G1 in Table \ref{table 1}. Here, a relatively weak correlation is present and one can expect that when using a diagonal scale matrix, the final adaptive values corresponding to $\mathbf{S}$ will be roughly the same for all three methods (given that $\Tilde{T} = \pi$ for MCT). 
The results in Table \ref{table 1} are very much in line with these predictions, and therefore the performance results are also very similar.

Next, a bivariate normal distribution with a larger difference in the marginal variances, represented by G2 in Table \ref{table 1}, is considered. Again, since the correlation is very weak in this example as well, one would expect a consistent behavior as in the former situation. 
Table \ref{table 1} indicates that all methods can handle such large differences in scale. Interestingly, the standard deviation across the estimates of the diagonal elements of $\mathbf{S}$ seems to be slightly smaller for MCT compared to the other two in both examples.

Next, a bivariate normal distribution with high correlation, given as G3 in Table \ref{table 1}, is considered. Due to the high correlation, it is not expected that all three methods yield the same outcome. Especially, one would expect significant differences between ISG and VARI, as the latter ignores the covariance. From Table \ref{table 1}, the mean of $\mathbf{S}$ from the ISG method differs significantly from the remaining two. For this distribution, the ISG method performs slightly better in terms of ESS per gradient evaluation. The MCT mean of $\mathbf{S}$ are close to the marginal standard deviations. Therefore, in multivariate normal distributions, it seems that the MCT will give results closer to the VARI method. This is not unreasonable as MCT  directly explores the dynamics of each coordinate individually, without directly considering the correlation between the coordinates. 

Finally, a higher dimensional setting, using a 10-dimensional standard normal distribution G4, is considered. Instead of checking all 10 coordinates, only the two coordinates that yield the smallest and largest ESS are presented in Table \ref{table 1}. 
It is seen that all methods produce reliable and expected results, even in a larger dimensional situation.

\subsection{Non-Gaussian distributions} \label{non-gaussian}

\subsubsection{NG1: Multivariate t-distribution} \label{t distribution}
In this example, a bivariate t-distribution  \citep[see e.g.][]{kotz2004multivariate} with $\nu = 4$ degrees of freedom is considered with the density given by:
\begin{equation} \label{4.1}
    \pi(\mathbf{q}) \propto \squarebrackets{
    1 + \frac{1}{\nu} (\mathbf{q} - \boldsymbol{\mu})^T \boldsymbol{\Sigma}^{-1} (\mathbf{q} - \boldsymbol{\mu})
    }^{-(\nu + 2)/2}, \, \boldsymbol{\mu} = (1,2)^T, \, \boldsymbol{\Sigma} = \begin{bmatrix}
    4 & 2 \\
    2 & 9
\end{bmatrix}.
\end{equation}
It can be shown that $E(\mathbf{q}) = \boldsymbol{\mu}$ if $\nu > 1$  and $\text{Var}(\mathbf{q}) = \frac{\nu}{\nu - 2} \boldsymbol{\Sigma}$ (if $ \nu > 2$). In particular, (\ref{4.1}) exhibits only a modest correlation of $1/3$. 

The former rows of Table \ref{table 2} present results obtained from the sampling procedure using the three techniques considered. In this case, ISG and MCT approach produce similar estimates of $\mathbf S$, whereas the VARI results are more variable, presumably related to the large variance of variance estimates in heavy tailed distributions. 
Regardless, Table \ref{table 2} shows no substantial differences in the number of ESS per gradient evaluation of the ODE system for all coordinates across the three methods. The increase in raw ESS for the VARI method gets weighted down as a consequence of a greater number of gradient evaluations due to the larger values of $\mathbf{S}$.

\begin{table}
\centering
\resizebox{\columnwidth}{!}{%
\begin{tabular}{ccccccccccc}
\toprule
 &  & \multicolumn{3}{c}{ESS} & \multicolumn{3}{c}{mean($\mathbf{S}$) (SD($\mathbf{S}$))} & \multicolumn{3}{c}{$\text{ESS} * 100000 $/ $N_{\text{ode}}$} \\
 \cmidrule(lr){3-5} \cmidrule(lr){6-8} \cmidrule(lr){9-11}
 & Method & VARI & ISG & MCT & VARI & ISG & MCT & VARI & ISG & MCT \\
 \cmidrule(lr){2-2} \cmidrule(lr){3-5} \cmidrule(lr){6-8} \cmidrule(lr){9-11}
Parameters & Position Coordinate &  &  &  &  &  &  \\
\midrule
\multirow{2}{*}{{NG1: Bivariate t-distribution}} & 1 & 423463 & 313629 & 318729 & 2.75 (0.212) & 2.17 (0.025) & 2.17 (0.055) & 1087 & 1035 & 1060 \\
 & 2 & 420223 & 320559 & 318059 & 4.17 (0.455) & 3.25 (0.038) & 3.23 (0.079) & 1079 & 1057 & 1058 \\
\cline{1-11}
\multirow{2}{*}{{NG2: Bivariate smiley distribution}} & 1 & 152750 & 123678 & 118629 & 0.98 (0.019) & 0.45 (0.021) & 0.99 (0.019) & 298 & 514 & 242 \\
 & 2 & 201172 & 130083 & 169444 & 1.66 (0.073) & 1.00 (0.019) & 1.34 (0.043) & 392 & 541 & 345 \\
\cline{1-11}
\multirow{2}{*}{{NG3: Bivariate bimodal distribution}} & 1 & 118129 & 52325 & 182349 & 0.91 (0.002) & 0.38 (0.011) & 2.04 (0.219) & 208 & 213 & 144 \\
 & 2 & 214633 & 105042 & 285056 & 1.35 (0.025) & 1.00 (0.019) & 1.43 (0.058) & 377 & 428 & 225 \\
\cline{1-11}
\bottomrule
\end{tabular}%
}
\caption{Results for the non-Gaussian distributions of Section \ref{non-gaussian}. $N_\text{ode}$ is the number of evaluations of the ODE system done by the numerical integrator in total over 10 independent trajectories. mean($\mathbf{S}$) and SD($\mathbf{S}$) correspond to the mean, and the standard deviation respectively, of the 10 values of the diagonal elements of $\mathbf{S}$ obtained from the 10 independent trajectories.}
\label{table 2}
\end{table}

\subsubsection{NG2: Smiley distribution} \label{smiley distribution}
Now, a case with a highly non-linear target distribution commonly known as a smiley distribution (see e.g. \cite{kleppe2022connecting}) is explored. The distribution considered here may be summarized by
\begin{align}
    q_2 \mid q_1 &\sim N(q^2_1, 1), \label{4.2} \\
    q_1 &\sim N(0, 1). \label{4.3}
\end{align}
From the middle rows of Table \ref{table 2}, one can observe that the methods behave differently from each other. In particular, the $S_1$ estimates obtained by VARI and MCT are very comparable (and consistent with what one would expect given the standard normal marginal of $q_1$), whereas the $S_2$ estimates differ considerably. The ISG tuning objective results in smaller values of $\mathbf{S}$ in contrast to the other two methods for both coordinates, 
with the procedure performing slightly better than the contenders. The usage of the gradient in the tuning objective of ISG seems to capture the non-linearity (and "locally varying correlation") of the distribution more efficiently than the remaining two. 

\subsubsection{NG3: Bimodal distribution} \label{bimodal distribution}
The final scenario that is examined in this group of distributions is a bivariate bimodal distribution represented by the following log density kernel:
\begin{equation}
    \log \Tilde{\pi}(\mathbf{q}) \propto -(1 - q_1^2)^2 - (q_2-q_1)^2 / 2.
\end{equation}
The distribution has two distinct modes and therefore imposes a rather complex structure. In particular, the local structure around each mode has smaller scales than what one would expect by simply looking at $\text{Var}(\mathbf q)$. 
Again, it is seen that different methods yield different estimates of $\mathbf{S}$ with MCT returning the largest scales. Presumably, this is related to that median crossings only occur for $q_1$ when the process jumps from one mode to the other.
The ISG approach again returns smaller values of $\mathbf{S}$ without sacrificing too many effective samples and is again the most efficient method (even if the difference relative to VARI is relatively small here). 

\begin{table}
\centering
\resizebox{\columnwidth}{!}{%
\begin{tabular}{ccccccccccc}
\toprule
 &  & \multicolumn{3}{c}{ESS} & \multicolumn{3}{c}{mean($\mathbf{S}$) (SD($\mathbf{S}$))} & \multicolumn{3}{c}{$\text{ESS} * 100000 $/ $N_{\text{ode}}$} \\
 \cmidrule(lr){3-5} \cmidrule(lr){6-8} \cmidrule(lr){9-11}
 & Method & VARI & ISG & MCT & VARI & ISG & MCT & VARI & ISG & MCT \\
 \cmidrule(lr){2-2} \cmidrule(lr){3-5} \cmidrule(lr){6-8} \cmidrule(lr){9-11}
Parameters & Position Coordinate &  &  &  &  &  &  \\
\midrule
\multirow{2}{*}{{F1: $q_2 \mid q_1 \sim N(0, \exp(1.5q_1))$}} & 1 & 216542 & 117841 & 130188 & 1.01 (0.023) & 0.67 (0.020) & 1.00 (0.045) & 368 & 447 & 344 \\
 & 2 & 137143 & 43171 & 42190 & 1.89 (0.190) & 0.56 (0.018) & 0.74 (0.023) & 233 & 164 & 111 \\
\cline{1-11}
\multirow{2}{*}{{F2: $q_2 \mid q_1 \sim N(0, \exp(2q_1))$}} & 1 & 43344 & 35832 & 46059 & 1.01 (0.070) & 0.59 (0.015) & 1.00 (0.033) & 46 & 139 & 104 \\
 & 2 & 1445 & 10778 & 5402 & 2.65 (0.785) & 0.37 (0.024) & 0.61 (0.029) & 2 & 42 & 12 \\
\cline{1-11}
\bottomrule
\end{tabular}%
}
\caption{Results for the funnel distributions. $N_\text{ode}$ is the number of evaluations of the ODE system done by the numerical integrator in total over 10 independent trajectories. mean($\mathbf{S}$) and SD($\mathbf{S}$) correspond to the mean, and the standard deviation respectively, of the 10 values of the diagonal elements of $\mathbf{S}$ obtained from the 10 independent trajectories.}
\label{table 3}
\end{table}

\subsection{Funnel distribution} \label{funnel distribution}
Finally, variants of a two-dimensional funnel distribution \citep[see e.g.][]{neal2003slice} are considered.
The different variants of the funnel distribution may be expressed as
\begin{align}
    q_2 \mid q_1 &\sim N(0, \exp(\omega q_1)), \label{4.4} \\
    q_1 &\sim N(0,1), \label{4.5}
\end{align}
where \cite{neal2003slice} (and several subsequent authors) considered $\omega=3$.
Choosing the rather extreme case of $\omega=3$ leads to highly variable/unreliable ESS estimates in some cases of the present setup, and instead the scenarios $\omega=1.5$ and $\omega=2$ are considered in Table \ref{table 3}. In addition to the presented results, it was verified that the simulated chains accurately represented the underlying standard Gaussian distribution of $q_1$. 


For the variant with $\omega=1.5$, none of the methods produce uniformly best sampling efficiency, even if there are quite large differences in the estimated $\mathbf S$, and also in raw ESSes.

When increasing $\omega$ to 2,
the performance of VARI drops substantially. Even though the mean estimates of $S_1$ from the MCT and VARI are comparable, the larger value of $S_2$ implies that the same amount of effective sample size is produced for the coordinate, but the number of gradient evaluations still increases.
For this specific choice of the funnel distribution, the ISG outperforms the other two. 

\section{Bayesian Logistic Regression } \label{bayesian logistic regression}
\begin{table}
\centering
\resizebox{\columnwidth}{!}{%
\begin{tabular}{lcccccccccc}
\toprule
 &  & \multicolumn{3}{c}{ESS} & \multicolumn{3}{c}{mean($\mathbf{S}$) (SD($\mathbf{S}$))} & \multicolumn{3}{c}{$\text{ESS} * 100000 $/ $N_{\text{ode}}$} \\
 \cmidrule(lr){3-5} \cmidrule(lr){6-8} \cmidrule(lr){9-11}
 & Method & VARI & ISG & MCT & VARI & ISG & MCT & VARI & ISG & MCT \\
 \cmidrule(lr){2-2} \cmidrule(lr){3-5} \cmidrule(lr){6-8} \cmidrule(lr){9-11}
Data set & Position Coordinate &  &  &  &  &  &  \\
\midrule
\multirow{2}{*}{{BLR1: Pima}} & max ESS & 1532594 & 755770 & 838729 & 0.17 (0.036) & 0.12 (0.005) & 0.13 (0.002) & 5453 & 3817 & 3301 \\
 & min ESS & 497898 & 293053 & 461428 & 0.16 (0.006) & 0.11 (0.003) & 0.15 (0.005) & 1771 & 1480 & 1816 \\
\cline{1-11}
\multirow{2}{*}{{BLR2: German credit}} & max ESS & 1614216 & 731038 & 884707 & 0.13 (0.037) & 0.08 (0.002) & 0.08 (0.001) & 3940 & 3510 & 2753 \\
 & min ESS & 386105 & 217883 & 335022 & 0.17 (0.024) & 0.08 (0.003) & 0.15 (0.008) & 942 & 1046 & 1043 \\
\bottomrule
\end{tabular}%
}
\caption{Results for the Bayesian logistic regression applied on the Pima and German credit dataset. $N_\text{ode}$ is the number of evaluations of the ODE system done by the numerical integrator in total over 10 independent trajectories. mean($\mathbf{S}$) and SD($\mathbf{S}$) correspond to the mean, and the standard deviation respectively, of the 10 values of the diagonal elements of $\mathbf{S}$ obtained from the 10 independent trajectories.}
\label{table 4}
\end{table}

As a real data application, we consider a Bayesian logistic regression model \citep[see e.g.][]{GelmanBDA3}. For the relatively large data sets considered, the posterior distributions are known to be quite close to Gaussian \citep{10.1214/16-STS581}, and hence we would expect that all $\mathbf S$-tuning methods should work satisfactory. 

The model is fitted to binary data $y_i\in \{0,1\}$, which are accompanied by covariate vectors $\mathbf x_i$, $i=1,\dots,n$. The model may be summarized by
\begin{align}
    y_i \mid \boldsymbol{\beta} &\sim \text{Bernoulli}(p_i), \, \log\parentheses{\frac{p_i}{1 - p_i}} = \mathbf{x}^T_i \boldsymbol{\beta}, \,  i = 1, 2, \dots, n, \label{4.6} \\
    \boldsymbol{\beta} &\sim N(\mathbf{0}_d, 100 \mathbf{I}_d). 
\end{align}
Two data sets are considered in this section: The Pima data \citep{ripley2007pattern}) ($n=532$, $d=8$ (including intercept parameter)) and the German credit data \citep[see e.g.][]{michie1994machine} ($n=1000$, $d=25$ (including intercept parameter)). Performance statistics for simulations targeting $p(\boldsymbol{\beta}|\mathbf y)$, using otherwise the same setting as in Section \ref{simulation}, are presented in Table \ref{table 4}.


For the Pima data, it is seen that MCT has the highest worst-case ESS per gradient evaluation. However, the differences across the methods are in terms of worst-case performance are not very large. This may in part be attributed to the fact that posterior is close to Gaussian, with no strong correlations (largest posterior correlation magnitude $\sim 0.6$).

For the German credit data, a similar pattern is observed, possibly with the VARI exhibiting slightly poorer worst-case performance than the contending methods. Still, the differences are small. For this data set, the posterior correlations are somewhat stronger, with some close to $0.8$ in magnitude, which may contribute to explain the relative deterioration of the VARI performance.

\section{Discussion} \label{discussion}


The present paper has presented and compared three different methods for tuning diagonal scale matrices for HMC-like methods. Though none of the methods are uniformly better than the remaining, it is certainly not clear that the common practice of simply using estimated marginal standard deviations for such scaling is always the best choice. In particular, in light of its easy implementation and inclusion in automatically tuned methods, the ISG method (including straightforward modifications for discrete time HMC implemented with reversible integrators) constitutes a worthwhile alternative to current practices, in particular when the target distribution exhibits strong correlations.

\bibliographystyle{chicago}
\bibliography{main}
\newpage

\appendix
\begin{center}
\Large Supplementary material to "Tuning diagonal scale matrices for HMC" by Jimmy Huy Tran and Tore Selland Kleppe
\end{center}
\setcounter{page}{1}
\section{Implementation of median crossing times} \label{implementation median crossing}
In Section \ref{median crossing}, the tools needed for the process of tuning the scale matrix $\mathbf{S}$ using median crossing times are introduced. Now, a summary of how the method is implemented will be presented.

First, it is mentioned that the median cannot be estimated continuously in a similar fashion as VARI and ISG using (\ref{2.9}). Under the setting of GRHMC, it is convenient to start out with $\mathbf{m}$ and $\mathbf{S}$ as the zero vector and identity matrix, respectively. A small part of the burn-in period, say $c_\text{burn} t_\text{burn}, \, c_\text{burn} \in [0, 1]$, is  spent specifically to build up a first representative estimate of $\mathbf{m}$ before the optimization of $\mathbf{S}$ takes place. By defining a grid of time points with equal spacing $\Delta$ over the burn-in period, i.e. $\Delta i, \, i = 1, 2, \dots, i_\text{burn}^*, \dots, i_\text{burn}$ such that $\Delta i_\text{burn}^* \leq c_\text{burn}t_\text{burn}$ and $\Delta i_\text{burn} \leq t_\text{burn}$, $\mathbf{m}$ is recalculated at step $i$ based on $\mathbf{q}(\Delta i)$ using the $P^2$-algorithm by \cite{jain1985p2}, a technique for estimating quantiles dynamically without the need of storing all previous observations. For high-dimensional problems and longer burn-in periods, this approach will significantly reduce the space complexity issues that occur in such settings when the standard method of estimating quantiles is applied. The reader is referred to the article for further details regarding the $P^2$-algorithm. After $i_\text{burn}^*$ steps, $\mathbf{m}$ is assigned to the estimated median vector at this time point and the tuning of $\mathbf{S}$ can start until the burn-in period ends while still letting $\mathbf{m}$ change at the remaining time points $\Delta i_\text{burn}^*, \dots, \Delta i_\text{burn}$.

An important note regarding the approach of MCT is that it is much more sensitive because of the stochastic mechanism of GRHMC. Due to the momentum refresh event, it is possible to obtain either very small or very large median crossing times, which in turns affects the estimates from the dual averaging scheme. The former case could occur if a momentum refresh event takes place only a short period of time after a median crossing event and changes the direction of motion towards the median again. On the other hand, large median crossing times are obtained whenever a momentum refresh event happens and alters the velocity in the opposite direction when the coordinate is very close to its median. Therefore, in order to obtain more stable and comparable estimates of $S_j$ when running the same setup multiple times, one can either increase $\gamma_j$ or $t_\text{burn}$. The drawback with increasing $\gamma_j$ is that the shrinking effect will become more apparent, especially for small values of $t_\text{burn}$, and the estimates of $\log(S_j)$ will not be able to move far away from $\mu_j$ during this short time span. Increasing $t_\text{burn}$ requires of course more computational power, which one wants to avoid if possible. 

To balance the issues above without increasing the demand of computational effort, the following strategy regarding the MCT approach is proposed:
\begin{itemize}
    \item Instead of one single run with MCT, a two-step strategy of burn-in using MCT is applied. Let $t_{\text{burn}_1}$ and $t_{\text{burn}_2}$ denote the duration of the simulation in the first and second step of burn-in. The first burn-in period will be considerably shorter as the intention of this part is to give a better initial configuration of the second burn-in period. Unless noted, $t_{\text{burn}_1} / t_{\text{burn}_2} = 0.2$. The default duration of the longer burn-in period $t_{\text{burn}_2}$ is 5000 time units. The total burn-in duration is therefore $t_\text{burn} = t_{\text{burn}_1} + t_{\text{burn}_2} = 6000$ for the default setting. For both runs, $\kappa_j$ and $\kappa_j^{(0)}$ from the dual averaging algorithm are set to 0.75 and 10 for all coordinates, respectively, and $c_\text{burn}$ is set to 0.05. The desired median crossing time for all coordinates is equal to $\Tilde{T} = \pi$ by default, a constant momentum refresh event rate of $\lambda = 0.2$ is considered and $\Delta = 1$ in the $P^2$-algorithm. 
    \item For the first run, there is no prior information available, and $\mathbf{m}$ is therefore initialized as the zero vector and $\mathbf{S}$ as the identity matrix. This leads to the choice of $\mu_j = 0$ for all $j$, i.e. the estimates of $S_j$ are shrunken towards unity. The shrinkage parameter for the first run is chosen to be $\gamma_j = 10$. The resulting estimate of $S_j$ obtained should at least be closer to the ideal value if it is indeed significantly different from unity. Therefore, the final value of $S_j$ and $m_j$ from this first step of burn-in can be used to initialize the second and longer burn-in period, which in theory should lead to faster convergence. 
    \item For the second step of burn-in, the estimates from the previous step are used as starting point for $\mathbf{m}$, $\mathbf{S}$ and $\mu_j$ in the dual averaging method by letting $\mu_j = \log(S_j)$. However, since the first run tends to underestimate when the true value is larger than 1, it can be beneficial to consider $\mu_j = \alpha_j \log(S_j)$ with $\alpha_j>1$ to compensate for this fact. By default, $\alpha_j$ is set to 1.1 for all $j$. Finally, in contrast to the first step, $\gamma_j$ is set to a larger value of $25$ as the procedure has now received a head start when initializing it with the results from the first run. This should reduce the variation in the results for different iterations while still obtaining final estimates that are close to the true value without increasing the duration of the burn-in period.

\end{itemize}

\section{Final setup} \label{final setup}
In order to get a sense of how the performance varies between the approaches given a specific model/situation, the following strategy is adopted for all three methods when simulating samples in Section \ref{simulation} and \ref{bayesian logistic regression}:
\begin{itemize}
    \item The burn-in period is now split into two parts such that $t_\text{burn} = t_{\text{burn}_{\mathbf{S}}} + t_{\text{burn}_{\lambda}}$: The first period of $t_{\text{burn}_{\mathbf{S}}}$ time units deals with the tuning of $\mathbf{m}$ and $\mathbf{S}$ given a specified and fixed constant event rate $\lambda_{\text{initial}}$. After this is done, $\mathbf{m}$ and $\mathbf{S}$ are fixed using the estimates from the recent run and the remaining $t_{\text{burn}_{\lambda}}$ time units of the burn-in period to select a suitable value of $\lambda$. In practice, both $\mathbf{m}$, $\mathbf{S}$ and $\lambda$ are tuned simultaneously. However, in order to get a better understanding of the presented methods themselves, this splitting is easier for quantifying the marginal effect of each scale matrix tuning approach. For the experiments to come, $t_{\text{burn}_{\mathbf{S}}} = 6000$ (same time length mentioned in the implementation of MCT in Appendix \ref{implementation median crossing}) while $t_{\text{burn}_{\lambda}} = 5000$. In addition, $\lambda_{\text{initial}}$ is chosen to be $0.2$ as default if it is not mentioned otherwise. 
    \item Regarding the tuning process of the momentum refresh event rate $\lambda$, the methodology presented in Section 4.3.2 by \cite{kleppe2022connecting} is applied. The default value of the smoothing parameter in the exponential smoothing process is 0.99. 
    \item After $t_\text{burn}$, $\mathbf{m}$, $\mathbf{S}$ and $\lambda$ are completely fixed and $N$ samples is generated for the remaining $t_\text{sample} = T - t_\text{burn}$ time units with equal spacing. 
    \item As a final comment, $\mathbf{m}$ and $\mathbf{S}$ are also initialized as the zero vector and identity matrix by default for the ISG and VARI approach just like in Appendix \ref{implementation median crossing} for the median crossing time method.
\end{itemize}

\end{document}